\documentstyle[twoside,fleqn,espcrc2,epsf]{article}

\newcommand{\ds}{\displaystyle}
\newcommand{\B}{\mbox{\it BeppoSAX\ }}

\title{The broad band spectral properties of galactic X--ray binary
pulsars}

\author{D. Dal Fiume\address{
Istituto Tecnologie e Studio Radiazioni Extraterrestri (TeSRE), C.N.R., \\
\ via Gobetti 101, 40129 Bologna, Italy},
M. Orlandini$^{\rm a}$, F. Frontera$^{\rm a}$\thanks{Also Physics Dept.,
Ferrara University},
S. Del~Sordo\address{%
Istituto Fisica Cosmica e Applicazioni all'Informatica (IFCAI), C.N.R., \\
\ via La Malfa 153, 90146 Palermo, Italy},
S. Piraino$^{\rm b}$, A. Santangelo${\rm ^b}$, A. Segreto${\rm ^b}$,
T. Oosterbroek\address{%
Astrophysics Division, Space Science Department of ESA, ESTEC, \\
\ Keplerlaan 1, 2200 AG Noordwijk, The Netherlands} and
A.N. Parmar${\rm ^c}$}
       
\begin{document}

\begin{abstract}
\B observed several galactic binary X--ray pulsars during the Science
Verification Phase and in the first year of the regular program. The
complex emission spectra of these sources are an ideal target for the \B
instrumentation, that can measure the emission spectra in an unprecedented 
broad energy band.\\
Using this capability of \B a detailed observational work can be done on
the galactic X--ray pulsars. In particular the 0.1--200 keV energy band
allows the shape of the continuum emission to be tightly constrained. A better
determination of the underlying continuum allows an easier detection of
features superimposed onto it, both at low energy (Fe K and L, Ne lines)
and at high energies (cyclotron features).\\
We report on the spectral properties of a sample of X--ray pulsars 
observed with \B comparing the obtained results.\\
Some ideas of common properties are also discussed and compared with our
present understanding of the emission mechanisms and processes.
\end{abstract}

% typeset front matter (including abstract)
\maketitle

\section{Introduction}

The instrumentation aboard \B \cite{sax,lecs,mecs,hp,pds} is 
particularly well suited to study the
X--ray emission from X--ray pulsars. This class of sources is composed
by binary systems in which a magnetized rotating neutron star accretes
matter from a less evolved mass--donor star. The mass--donor may be
a OB supergiant star as in the case of Vela X--1, a Be
main--sequence or near--main--sequence star as in the case of transient
recurrent pulsars like A0535+26, a low mass star as in the case of
4U1626--67. The type of mass donor star strongly affects the temporal
behaviour on the medium (days) to long (years) time scales. The
transient behaviour is almost completely restricted to the subclass of
X--ray pulsars that have Oe or Be counterparts.

\B has observed some persistent pulsars and one transient pulsar during
the first year of its operative life. We report results from the
observations of some of these sources, emphasizing the commonalities and
the differences. In particular we discuss the observational evidence on
cyclotron line feature, comparing the observed results, also in terms of
possible correlations, with the expected ones on the basis of the
available theoretical models.

\section{Observations}

X--ray pulsars are a relevant section of the \B Core Program
\cite{perola} devoted to compact galactic sources. During the 
Science Verification Program (SVP) and the AO1 regular program a number
of sources in this class were observed. In this paper we report the
results on Her X--1 (SVP), Vela X--1 (SVP), 4U1626--67 (SVP), Cen X--3
(AO1) and GS1843+00 (AO1).
The SVP observaton of Cen X--3 will not be discussed here.
A log of the \B observations is reported in Table 1

\begin{table*}
\label{tab:log}
\caption{\B log of observations}
\begin{tabular}{|l|l|l|}
\hline
                 & & \\
  Source Name    & Observation Date & Live Time on Source \\
                 & & (seconds) \\
\hline
Her X--1         &  1996/07/24--28 (SVP)     & 90000  \\
Vela X--1        &  1996/07/14 (SVP)         & 21600  \\
4U1626--67       &  1996/08/06 09--10 (SVP)  & 97000  \\
Cen X--3         &  1997/02/27--28 (AO1)     & 20000  \\
GS1843+00        &  1997/04/04 (AO1)         & 22000  \\
\hline
\end{tabular}
\end{table*}

Details of the single observations can be found elsewhere (Her X--1:
\cite{herx1}; Vela X--1: \cite{velax1}; 4U1626--67: \cite{1626}; 
Cen X--3 \cite{cen1,cen2}; GS1843+00 \cite{1843}).

All the observations we report were performed with all \B telescopes,
covering the energy band 0.1--200 keV. The observations were performed
before the failure of the MECS 1 HV module occurred in May 1997, 
therefore all the three MECS telescopes are available.
The live time in column three is calculated for the MECS telescopes. The
LECS live time is $\sim 40\%$ of that due to the fact that the instrument
is operated only during satellite night time. The HPGSPC and PDS live
time is $\sim 50\%$ of that, due to the background measure with the
rocking collimators.

In all the observations but for Vela X--1 the rocking instruments 
(HPGSPC and PDS) were
using the default collimator laws: one rocking every 96s.
In the Vela X--1 observation the dwell time was 50s.
All data were transmitted using the default direct telemetry modes,
apart from the observation of Her X--1, in which the HPGSPC data were
transmitted using the full diagnostic mode.

Here we report and compare the results on the pulse--phase averaged
spectra in 0.1--200 keV.
In four of these five sources, we have evidence of the presence of a
cyclotron line.

\section{Results}

The modeling of the broad--band spectra of X--ray pulsars is quite
difficult. The Her X--1 spectrum is a prototype of this complexity
\cite{herx1}.
It needs several different spectral components to be reasonably modeled:
a low energy cutoff,
a low energy black body -- likely due to reprocessing of primary X--ray
radiation at the magnetospheric boundary--, a Fe L line, a Fe K line,
a power law, a high energy exponential cutoff and a cyclotron feature.
A formal fit to the observed data has 15--19 free parameters. Only the
\B capability to observe the energy spectrum in more than four decades
in energy allows the clear separation of the different components resulting
in formal fits in which the correlations between parameters are reasonably
low.

Currently there is no theoretical model that provides a parametrized
model function to fit the observed spectra. Therefore the fit to the
observed count rate spectra can be performed only using {\it ad hoc}
analytical models. The first and more widely used model function to
describe the broad band continuum is the
power--law--plus--cutoff \cite{whish} that gives a quite reasonable
empirical description of the shape of the energy spectra of X--ray
pulsars.
\begin{equation}
\begin{array}{ll}
E^{-\alpha} & E<E_{cutoff}\\
E^{-\alpha}\times \exp(-\frac{\ds E-E_{cutoff}}{\ds E_{folding}}) 
& E\stackrel{>}{=}E_{cutoff}
\end{array}
\end{equation}
This model function has a discontinuity at E=E$_{cutoff}$.

\begin{table*}
\caption{Cyclotron lines in the spectra of X--ray pulsars}
\label{tab:fit}
\begin{tabular}{|l|l|l|}
\hline
  Source Name & ECycl Feature   & FWHM Cycl. \\
              & keV$^{(a)}$     & keV  $^{(a)}$ \\
\hline
 & & \\
Cen X--3  & 28.5 $\pm$ 0.5 & 6.3$\pm$2.0 \\
4U1626--67 & 36.5$\pm$1.0  & 7$\pm$2.8 $^{(b)}$  \\
Her X--1   & 42.1$\pm$0.3  & 14.7$\pm$0.9  \\
Vela X--1  & 57.9$\pm$1.0  & 24.0$\pm$1.  \\
\it A0535+26   & \it 110.$^{(c)}$  & \it 56$^{(c)}$ \rm \\
\hline
\multicolumn{3}{l}{\footnotesize (a) Using a multiplicative Gaussian 
absorption function}\\
\multicolumn{3}{l}{\footnotesize (b) The FWHM is significantly larger
using a Lorenzian function \cite{1626}}\\
\multicolumn{3}{l}{\footnotesize (c) Using a Lorenzian function 
\cite{grove}}
\end{tabular}
\end{table*}

Other widely used functions are the Fermi--Dirac CutOff (FDCO --
\cite{tanaka})
\begin{equation}
E^{-\alpha}  \frac{1}{1+\exp(\frac{\ds E-E_{cutoff}}{\ds E_{folding}})}
\end{equation}
and the Mihara Negative and Positive power--law plus EXponential
(NPEX -- \cite{miharath})
\begin{equation}
(A_1 E^{-\alpha_1} + A_2 E^{+\alpha_2}) \times \exp(-\frac{E}{kT})
\end{equation}
We tried all the model functions listed above for the sources in  Table
\ref{tab:log}. All these model functions show the same practical problem:
they do not adequately describe the shape and the bending of the cutoff,
especially in the energy interval where the power law joins the
exponential tail.

We also tried for all sources a ``modified'' power--law--plus--cutoff
model, in which we used a broken power law:
\begin{equation}
\label{eq:bknpo}
\begin{array}{ll}
E^{-\alpha_1} & E<E_{break} \\
E^{-\alpha_2} & E\stackrel{>}{=}E_{break} \\
E^{-\alpha_2}\times \exp(-\frac{\ds E-E_{cutoff}}{\ds E_{folding}}) 
& E\stackrel{>}{=}E_{cutoff}
\end{array}
\end{equation}

This empirical function seems to provide the best performance in
fitting the count rate spectra {\it on the average}, therefore it can be
used to compare the results from the different sources.

In Table \ref{tab:fit} we summarize the results of the broad band fits
concerning the cyclotron line. The values reported in this table come
from fits using the model functions in equation \ref{eq:bknpo} to describe
the broad band continuum. The results for A0535+26 are taken from Grove
et al. \cite{grove}.
In this table all measurements apart that of A0535+26 were done with
\B.

In GS1843+00, the source in our sample with the hardest spectrum
\cite{1843}, we do not detect any cyclotron feature in the pulse--phase
averaged spectrum.

The results in this table suggest that there may be a correlation
between the cyclotron feature centroid energy and its FWHM. 
This correlation is reported in Figure 1, where we also
added the data on the cyclotron feature observed in A0535+26 by OSSE
\cite{grove}.
We stress that the FWHM parameter in the fit to the count rate spectra
shows a possible dependence on the choice of the function modeling the
broad band continuum, therefore this correlation must be taken with
caution. We tried to limit this effect due to the choice of the
continuum model function, using the same broad band continuum in all fits.
The data follow a general linear trend (with slope $\alpha \simeq 1.7$)
with some evident scatter.

We also caution that the line energy reported here may not be the energy
of the fundamental cyclotron feature, but its first harmonic. This may
be the case of Vela X--1 and A0535+26.

A summary of the cyclotron lines measured by \B is shown in Figure 2. In
this Figure we plot a ``modified Crab ratio'' as described in
\cite{herx1,velax1}. To obtain this ratio we
\begin{itemize}
\item divided the count rate spectrum by the measured Crab count rate
spectrum
\item multiplied the result by the slope of the Crab Nebula photon
spectrum (E$^{-2.1}$)
\item divided the result by the continuum model function used for the broad
band fit
\end{itemize}
This procedure enhances the deviations from the broad band continuum and
gives a way to \underline{\it visually} compare the different features
observed in the different sources.

\section{Discussion}

The correlation showed in Figure 1 was ``qualitatively'' predicted by
M\'esz\'aros and Nagel \cite{mn} (see also \cite{pulmod}).
The model predicts a width of the cyclotron feature proportional to its
energy and to the square root of the electron temperature of the atmosphere
\begin{equation}
\label{eq:fw}
\Delta \omega_B \simeq \omega_B \left( 8 \times \ln(2) \times 
\frac{\rm kT_e}{\rm m_ec^2} \right)^{\frac{1}{2}} |\cos\theta|
\label{eq13}
\end{equation}
In this equation $\Delta\omega_B$ is the line width, $\omega_B$ is the
cyclotron line frequency, $T_e$ is the electron temperature and
$\theta$ is the viewing angle with respect to the magnetic field axis.
A better insight on the properties of the cyclotron lines can be
obtained with pulse--phase resolved spectroscopy, as equation 
\ref{eq:fw} suggests that there may be a dependence on the viewing angle
of the observed line width (see also \cite{pphspe}).

However Araya and Harding (1996) \cite{araya} caution that, in the limit
of a single scattering, the line width is not related to the electron
temperature.

This ambiguity in the interpretation of these observational data points
out the need of a more detailed and quantitative model for the line
properties and for the broad band continuum emission of X--ray pulsars

\clearpage

\begin{figure*}
\epsfxsize=17cm
\centerline{\epsfbox{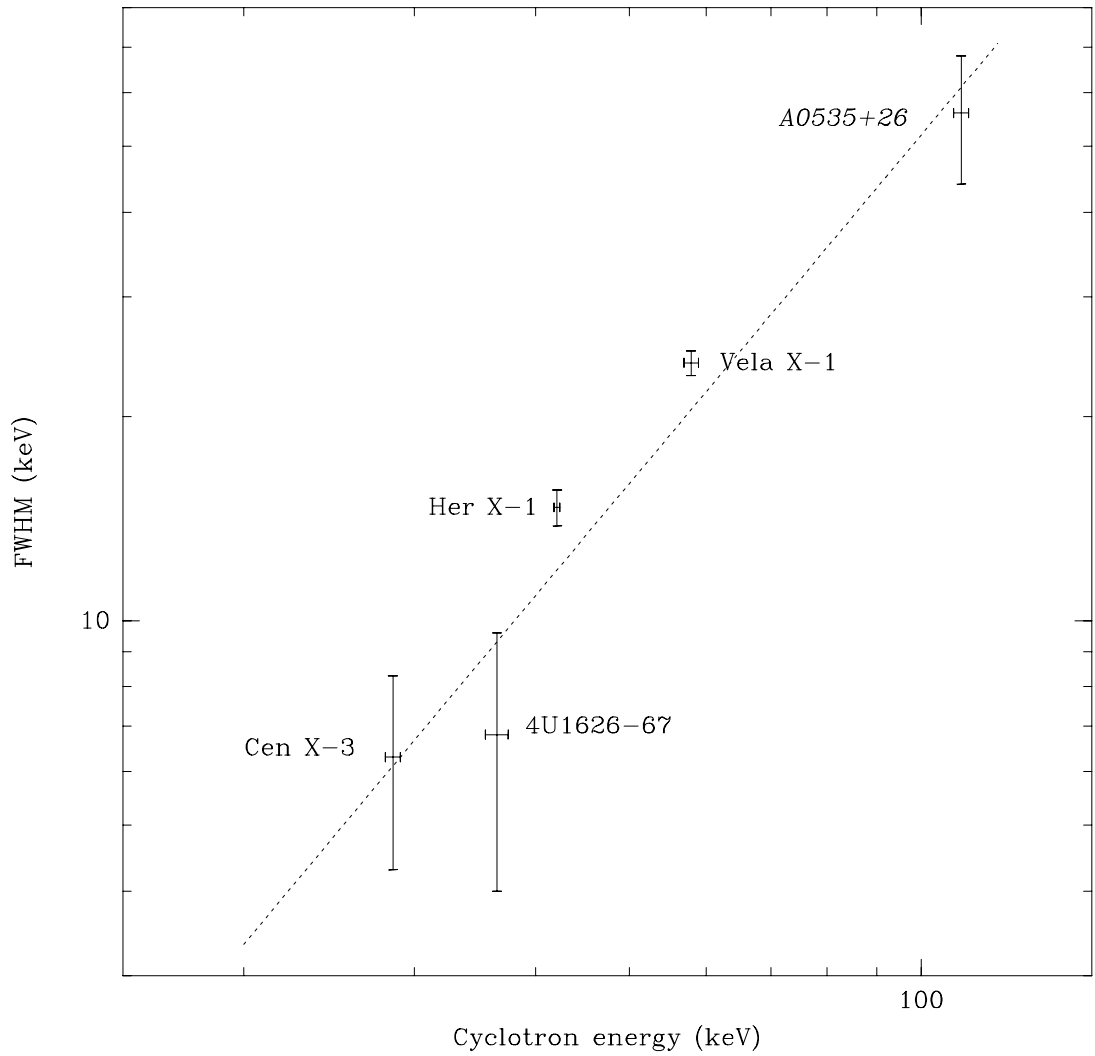}}
\caption{Cyclotron feature FWHM versus centroid energy. The dashed line
shows a simple least square linear fit. Even with some scatter present
in data (see text for a discussion), a correlation between these
quantities seems present.}
\end{figure*}

\clearpage

%\epsfxsize=17cm
%\epsfysize=10cm
%\centerline{\epsfbox[150 150 400 400]{all_cyc.ps}}
\begin{figure*}
\vspace{10cm}
\centerline{\includegraphics{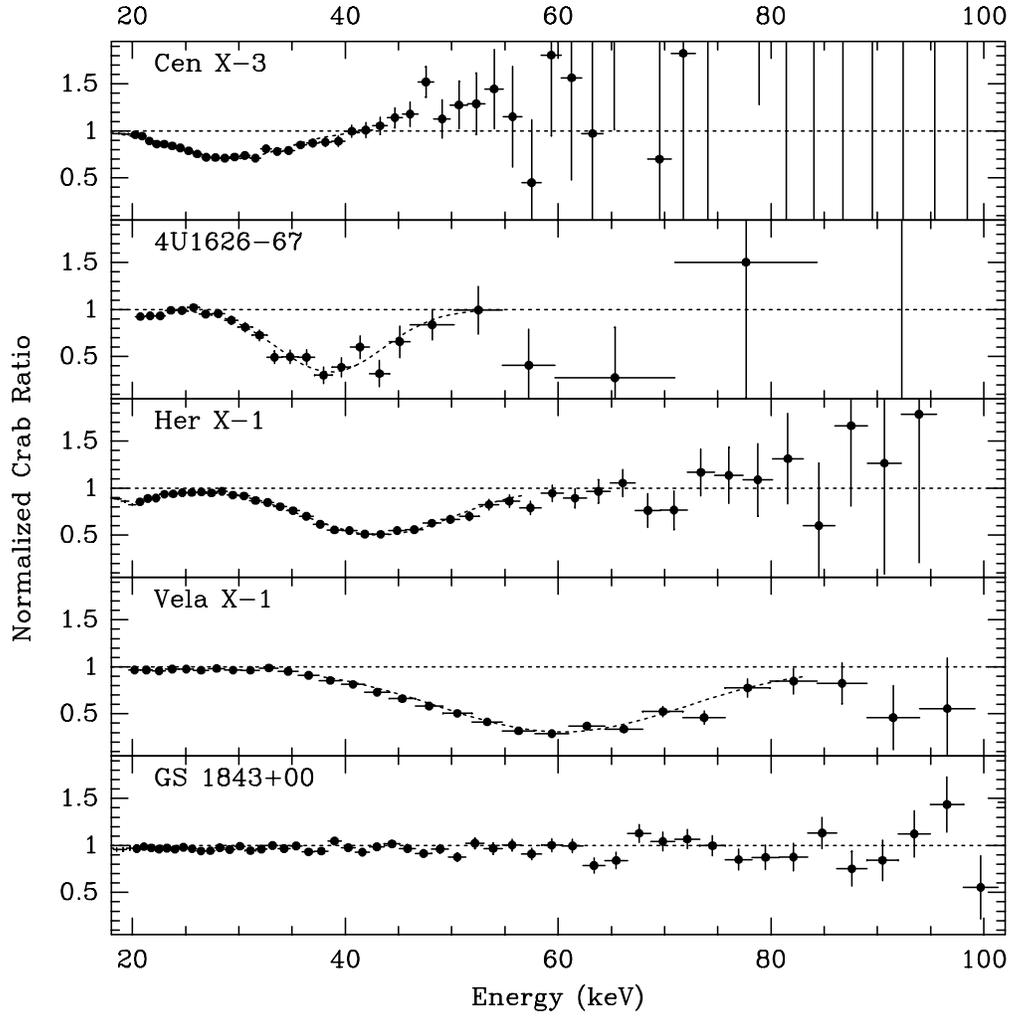}}
\caption{Ratios between the count rate spectrum of the different sources 
and the Crab count rate spectrum. The ratios are
normalized to the shape of the broad band continuum (see text for
details). The shape and position of the cyclotron features are well
evident. The spectrum of GS1843+00 does not show any deviation from the
shape of the broad band continuum}
\end{figure*}

\end{document}